\documentclass{article}
\usepackage{epsf,graphicx}

\newlength{\refindent}
\newlength{\parskiplen}
\begin{document}


\newenvironment{references}{\clearpage
			    \section*{\large \bf REFERENCES}
			    \parindent=0mm \everypar{\hangindent=3pc
			    \hangafter=1}}{\parindent=\refindent \clearpage}
\newenvironment{figcaps}{\clearpage
			 \section*{\large  \bf FIGURE CAPTIONS}}{}
\newcommand{\fig}[2]{\parbox[t]{2.0cm}{Figure #1:} \
		   \parbox[t]{13.5cm}{#2}\\[\baselinestretch\parskiplen]}


\begin{titlepage}
\begin{center}
\vspace*{0.5cm}
{\huge Photometry of the Magnetic White Dwarf }\\[0.3cm]
{\huge SDSS~121209.31+013627.7}\\[3.0cm]

{\large C. Koen$^1$ and P.F.L. Maxted$^2$}\\[1cm]
\normalsize
{\em 1 Department of Statistics, University of the Western Cape,
Private Bag X17, Bellville, 7535 Cape, South Africa}\\[0.2cm]
{\em 2 Astrophysics Group, Keele University, Keele,
      Staffordshire ST5 5BG, United Kingdom}  \\[2.0cm]

\end{center}

\begin{quotation}\noindent{\bf ABSTRACT.}
The results of 27 hours of time series photometry of 
SDSS~121209.31+013627.7 are presented. The binary period established
from spectroscopy is confirmed and refined to 0.061412 d 
(88.43 minutes). The photometric variations are dominated
by a brightening of about 16 mmag, lasting a little less
than half a binary cycle. The amplitude
is approximately the same in $V$, $R$ and white light. A secondary
small brightness increase during each cycle may also be present.
We speculate that SDSS~121209.31+013627.7 may be a polar in a low
state.

\vspace*{1.0cm}
\end{quotation}

\end{titlepage}

\section{INTRODUCTION}

The white dwarf star SDSS~121209.31+013627.7 (abbreviated ``SDSS~1212+0136" 
below) has a mean surface magnetic field $\sim 7$ MG, derived from
Zeeman splitting of its Hydrogen absorption lines (Schmidt et al. 2005a).
Spectra also show weak H$\alpha$ emission, implying a composite system.
Despite the relatively low temperature of the white dwarf ($\sim 10000$ K)
there are no overt signs of the presence
of the companion in photometry blueward of the $J$ band. This suggests that
the companion is very cool -- probably a brown dwarf (Schmidt et al. 2005a).
Radial velocity changes and modulation of the H$\alpha$ equivalent width
allow a binary period of about 90 minutes to be deduced. The orbital 
inclination appears to be high.

Given the close proximity of the two components (implied by the short binary 
period), the cool companion is strongly irradiated by the white dwarf. The
H$\alpha$ emission may then be ascribed to re-radiation from the facing 
hemisphere of the cool object.

Since radiation from the white dwarf is dominant even at $J$, Schmidt et al.
(2005a) suggested follow-up photometry and spectroscopy further into the 
infrared,
at $K$. The authors also speculated that optical photometry may be useful to
test for pulsation in the white dwarf, and to search for possible eclipses
of the white dwarf by the unseen companion. The latter two aims motivated
the observations reported here.

The experimental work is described in section 2, and the analysis of the
data in section 3. Section 4 deals with the modelling of the photometry,
and conclusions are presented in section 5.
 
\section{THE OBSERVATIONS}

All measurements were made with the SAAO (South African Astronomical
Observatory) CCD camera mounted on the SAAO 1.9-m telescope at
Sutherland, South Africa. The camera, which has a field of view 
of about 2.5 arcmin $\times$ 2.5 arcmin, was operated in $2\times 2$ prebinning
mode, which gave a reasonable readout time of 20 seconds. A log of the
observations is given in Table 1. Measurements comprising the last three runs
were made in white light, i.e. no filter was placed in the light beam;
this allowed shorter exposure times to be used, and hence better
time resolution to be obtained. The effective wavelength of these white-light
observations is between $B$ and $V$, but with a very wide bandpass. 
Photometric reductions were performed
using an automated version of {\textsc DOPHOT} (Schechter, Mateo \& Saha
1993).

Typically only four measurable stars were visible in the field of view.
The brightest two stars were used to differentially correct the photometry
for atmospheric effects. The remaining two stars were of comparable
brightness, and the non-programme star is therefore a useful ``check" star
for the photometry of SDSS~1212+0136. The nightly mean differences
[in the sense (check star)-(programme star)] are shown in Table 1.
There is some evidence for changes in the mean light level of
SDSS~1212+0136 from night to night. The last column of the Table gives
the standard deviations of the photometry of the check star: since its
mean magnitude is quite similar to that of the programme star, these values
probably constitute reasonable estimates of the photometric accuracies
of the measurements of SDSS~1212+0136. Standard deviations for the two
local comparison star measurements were typically half (filter-less 
observations) or one third ($V$-band) of those in the Table.

The results of the three white-light runs are plotted in Fig. 1. The
data have been adjusted to have the same nightly mean values. Bumps in
the light curves, roughly 0.05 d apart, are clearly visible. It is noteworthy
that the scatter in the three lightcurves is not correlated with the quality
of the nights, as measured by the $\sigma$-values in Table 1, but rather
with the mean light level of the observations ($\Delta z$-column in Table 1).
The correlation is in the sense that the brighter SDSS~1212+0136, 
the more noisy its white light lightcurve.

The $V$-band lightcurves (Fig. 2) are generally more noisy than those obtained
without any filter: this is probably mainly due to the fact that the brightness
in $V$ is substantially ($\sim 1.5$ mag) fainter than in white light. The
reader's attention is drawn to the small brightness increases about
midway between the bumps in the light curves: see particularly the
lightcurves for HJD~2453824 and 2452826. Fig. 3 compares results in $R$, $V$
and white light: it is interesting that the sizes of the bumps in the light
curves are comparable despite being observed through different filters.

\section{DATA ANALYSIS}

Fig. 4 shows the amplitude spectra of all the $V$ data (top panel)
and all the data acquired without any filter (bottom panel). 
(In order to avoid confusion we mention that the term ``amplitude"
should be understood to mean ``semi-amplitude"; we will refer explicitly 
to ``peak-to-peak amplitude" when that meaning is intended).
A telling feature
of the two spectra is excesses of power around frequencies 8, 16 and 
32 d$^{-1}$. Schmidt et al. (2005a) determined a period of 0.065 d
(frequency $f=15.4$ d$^{-1}$) for the star: this evidently corresponds to the 
main peak
at about 16 d$^{-1}$, with the other two frequencies being respectively
the first harmonic and a sub-harmonic. The sub-harmonics are evidently
induced by cycle-to-cycle variations in the observed lightcurves, and
will be ignored in what follows: the statistical model
\begin{eqnarray}
m(t)&=&\sum_{j=1}^2 C_j \cos(2 \pi jft+\phi_j)+e_t\nonumber\\
 &=& \sum_{j=1}^2 [A_j \cos (2\pi jft)+B_j \sin (2\pi j ft)]+e_t
\end{eqnarray}
is fitted to the data $m(t)$. In (1); $C_1$, $C_2$ and $\phi_1$, $\phi_2$
are the amplitudes and phases associated with the fundamental frequency
$f$ and its first harmonic $2f$; and $e_t$ is an error term. Expansion of 
the cosine form into the sum of a cosine
and a sine [the second form in (1)] leaves the frequency $f$ as the only
nonlinear parameter in a regression of the observations $m_t$ on time $t$.

The results of fitting the model (1) to the $V$-data and filter-less
observations can be seen in Fig. 5, for a range of trial frequencies $f$.
Note that for a given value of $f$, the parameters $\mu$, $A_1$, $A_2$
$B_1$ and $B_2$ all follow from minimisation of the sum of squares
\begin{equation}
SS(f)=\sum_t \left \{ m(t)-
 \sum_{j=1}^2 [A_j \cos (2\pi jft)+B_j \sin (2\pi j ft)] \right \}^2 \; .
\end{equation}
The logarithm of the minimised $SS(f)$, denoted $RSS(f)$ (``residual sum of 
squares") is plotted in the figure. 

Fig. 6, which is a more detailed view of the
frequency interval of principal interest, shows very good agreement between
the $V$ and white light results. Of course, the frequency resolution in the
latter is considerably better, since the observations cover
a longer time baseline (24 versus 3 days). It is particularly useful that the
alias peaks near 15.8 and 17.8 d$^{-1}$ in the bottom panel of the diagram
can be discounted since there are no substantial counterparts in the top panel.

The best-fitting frequencies $f_*$ are 16.288 d$^{-1}$ ($V$ data) and
16.2836 d$^{-1}$ (white light).
The interpretation of Fig. 6 is aided by noting that the quantity
\begin{equation}
\Lambda=(N-p) \left [ \frac{\sum_t RSS (f)}{\sum_t RSS(f_*)}-1 \right ],
\end{equation}
which is the approximate Gaussian likelihood ratio, has an approximate
$\chi^2_1$ distribution (i.e. chi-squared with one degree of freedom --
see Gallant 1987 for the theory, and Koen 2004 for an application very 
similar to the present one). In (3), $N$ is the number of observations, and $p$
is the number of estimated parameters. In Eqn. (2), the parameters $A_1$, $A_2$, $B_1$,
$B_2$ and $f$ are estimated; in addition, the nightly offsets from a common mean value
were calculated. It follows that $p=9$ for the $V$ data, $p=8$ for the white light
observations.

Koen (2004) showed that for 
large $N$ (a few hundred or more), (3) implies that $\alpha$-level confidence
intervals for $f$ are given by
\begin{equation}
\Delta \log RSS=
\log RSS(f)-\log RSS(f_*)  \stackrel{<}{~} \frac{\chi^2_1(\alpha)}{N-p} \; .
\end{equation}
In the present case the number of observations is 394 ($V$) or 526 (white 
light), hence $\Delta \log RSS \approx 0.0100, 0.0074$ respectively, for
95\% confidence intervals [$\chi^2_1(0.95)=3.84$]. The implication is that
the aliases are comfortably outside the 95\% confidence interval for the best
frequency.

Details of the best fitting solutions for the two datasets appear in 
Table 2. Also shown are the results of fitting the model in (1) to
the $R$-band data using $f=16.2836$ d$^{-1}$. Fig. 7 shows the data 
folded with respect to the best-fitting period, and amplitude
spectra of the residuals are plotted in Fig. 8. It is interesting that
the model has accounted for most of the $V$-band power at the sub-harmonic,
but little of the power in the sub-harmonic of white light observations.
We ascribe this to greater variability in the shape and/or mean level
of the filter-less measurements.

Since the folded lightcurves are noisy their shapes are extracted by 
statistical means. First, the fits of Eqn. (1) provide the top curves
in each of the three panels in Fig. 9. The small bumps near phase 0.5 
referred to at the end of section 2 clearly form part of 
that simple parametric model. The remaining two curves in each panel were
obtained by a non-parametric smoothing method known as ``loess" (Cleveland
\& Devlin 1988; Cleveland, Devlin \& Grosse 1988). The technique is akin
to a weighted moving average, but can be applied to irregularly spaced
data. In our implementation, we used local quadratic fits to the phased
data in Fig. 7, and to the similarly phased $R$-band data. Window widths
of 0.4 (middle curves) and 0.3 (bottom curves) of a cycle were used for the 
abundant $V$ and white-light data; for the sparser $R$ data the window
widths were 0.5 and 0.4 of a cycle.

The smaller the window width used the more detail emerges, at the risk
of extracting spurious feature due to noise in the data. The following 
generalities nonetheless seem plausible:
\begin{itemize}
\item[(1)]
The lesser bump in the light curve seems to be a real feature, and may
increase in amplitude towards longer wavelengths.
\item[(2)]
Small bumps aside, there appears to be a systematic decrease in brightness
between the end of one large bump and the onset of the next.
\item[(3)]
There is an impression that the large bumps may be flat-topped in $V$ and $R$,
but not in white light.
\item[(4)]
The large bumps appear symmetrical in $V$ and $R$, but not in white light.
\end{itemize}

The next section of the paper deals with attempts to explain the light 
curve shapes.

\section{PHYSICAL MODELS}

 We assume that the main periodicity in our lightcurves is the same as the
orbital period of SDSS~1212+0136, given that it is very similar to the period
seen in the radial velocity of the weak H$\alpha$ emission line by Schmidt
et~al (2005a). We used the four radial velocities presented in their fig.~3 
for the
dates listed in their fig.~2 to re-calculate the semi-amplitude of the orbit,
$K$, using a least-squares fit of a cosine with the data equally weighted and
with the orbital period fixed at the value derived above. Other free
parameters in the fit were the time of maximum positive radial velocity, $T_0$
and the systemic velocity, $\gamma$. We account for the finite exposure time
of about 500s in our fitting procedure and find that this increases the
value of $K$ by about 8\,km\,s$^{-1}$. We find $K=355\pm 6$\,km\,s$^{-1}$,
$\gamma=33 \pm 4$\,km\,s$^{-1}$ and $T_0=0.2801 \pm 0.0003$\,days, where
$T_0=0$ corresponds to 0000UT 2005 May 14. With this value of $K$ we find that
the companion is close to filling its Roche lobe. For example, for a companion
with a mass of $0.05~M_\odot$ the radius of the Roche lobe is 0.11$R_\odot$,
which is comparable to the radius expected for a typical brown dwarf 
(Baraffe et al. 2003).

 It is reasonable to assume that the variability in the lightcurves is the
reflection effect caused by the irradiation of one side of the companion by
the white dwarf. We can estimate the amplitude of the reflection effect in
magnitudes, $\Delta m_{\rm ref}$, using
\[\Delta m_{\rm ref} \approx -2.5 \log \left( \frac{f_{\rm opt,wd} + f_{\rm int}
f_{\rm opt,ref}}{f_{\rm opt,wd}}\right),\]
where $f_{\rm int} \approx 0.01$ is the fraction of the white dwarf's light
intercepted by the companion, $f_{\rm opt,wd}$ is the fraction of the white
dwarf's light emitted at wavelengths covered by our white light, V-band and
R-band photometry and $f_{\rm opt,ref}$ is the fraction of the intercepted
flux re-emitted over the same wavelength region. We used the pure hydrogen
model atmosphere spectra for a 10,000\,K white dwarf by Rohrmann (2001)
to estimate $f_{\rm opt,wd} \approx 0.36$. If the
irradiated hemisphere of the companion re-emits all the intercepted flux and
we assume an effective temperature for the companion of about 1600\,K, then
the net flux from this irradiated hemisphere is equivalent to an effective
temperature of 1900\,K. For blackbody radiation, this gives $f_{\rm opt,ref}
<0.01$, which would imply that the reflection effect would be undetectable at
optical wavelengths. More realistically, the re-emitted spectrum is likely to
be dominated by radiation from the point on the companion closest to the white
dwarf, which will be hotter than the average temperature on the irradiated
hemisphere. Even so, the maximum value of  $f_{\rm opt,ref}$ for blackbody
radiation, which occurs for temperatures of about 8500\,K, is $f_{\rm
opt,ref}\approx 0.45$. For this extreme value we obtain a peak-to-peak
amplitude $\Delta m_{\rm ref} =0.014$, less than half the observed amplitude.
We conclude that the main cause of the photometric variability in 
SDSS 1212+0136 is not the reflection effect, but that the reflection effect 
may be a small contribution to the variability, .e.g., it may be the cause of 
the small bump seen in the lightcurve.

Similar arguments can be applied to the H$\alpha$ emission line, as has been
done by Schmidt et~al. (1995) for the white dwarf -- M-dwarf binary star 
GD~245. In the case of this emission line we
use the flux of ionizing photons incident on the brown dwarf surface to
calculate an upper limit to the apparent flux in the H$\alpha$ emission line
by assuming that one
 H$\alpha$ photon is emitted for every incident photon below
the Lyman limit and that there are no other sources of H$\alpha$ emission. The
equivalent width of the H$\alpha$ line is then given by \[ {\rm EW}(\rm
H\alpha) \approx f_{\rm int}\cdot Q =  f_{\rm int}\frac{\int_0^{912{\rm
\AA}}S_{\lambda}d\lambda}{S_{\lambda}(H\alpha)},\] where $S_{\lambda}$ is the 
photon flux from the white dwarf. This can be compared directly to the value of
EW(H$\alpha$)$\approx 10$\AA\ presented by Schmidt et~al (1995). We used 
the pure
hydrogen model atmosphere of Rohrmann (2001)
for $T_{\rm eff}=$10,000\,K, $\log g=8$
to find $Q=0.33$. For this value of $Q$ the equivalent width of the H$\alpha$
is expected to be $<0.01$\AA.

 Given the large observed amplitude and the strength of the 
H$\alpha$ emission line, it appears that there is an additional source of
radiation in the system. Accretion onto the
white dwarf is a likely candidate. We can estimate
the accretion rate by assuming that all of the energy due to accretion
intercepted by the companion is emitted in the H$\alpha$ line. Using the
distance of 145 pc estimated by Schmidt et~al. (2005a), we find the accretion 
rate onto the white dwarf is about $10^{-13} M_\odot/y$. This is in good agreement with
the mass transfer rates onto white dwarfs from the solar-type wind 
of the low mass companion in other pre-polar binary systems (Schmidt et al. 2005b).
Assuming that there is accretion onto the magnetic white dwarf in this binary,
it is likely that cyclotron emission contributes to the optical variability we
have observed. We do not consider it worthwhile to discuss a model for
cyclotron emission in detail here given the large number of parameters
required for such a model and the small number of constraints that can be
imposed on these parameters given the existing observations.

If there is an accretion hot spot on the white
dwarf where material from the companion is channelled onto one of its magnetic
poles, then the change in brightness could be due to the change in visibility
of the spot as the white dwarf rotates. Occultation
as the spot disappears over the limb of the white dwarf, or 
non-isotropic emission from the spot, could explain the variability.  
A very approximate estimate of the
characteristic temperature of such a hot spot can be made by assuming that the
increase of about 5\,percent in brightness at the maximum of the lightcurve is
due to optically thick radiation from a spot with a radius of
5\,--\,10\,percent of the white dwarf radius powered by accretion at a rate
equivalent to 5\, percent of the luminosity of the white dwarf. 
In this case the
temperature is 15\,000\,--\,20\,000\,K. These high temperatures are consistent
with the observation that the increase in brightness is similar in the $V$ and 
$R$ bands, and the white-light photometry.

\section{CONCLUSIONS}
 
We confirm and refine the orbital period of about 90 minutes for
SDSS~1212+0136 seen by Schmidt et~al. (2005a) using lightcurves in various
optical bands. There are no obvious eclipses in our lightcurves of
SDSS~1212+0136, neither is there any sign of pulsation.

The lightcurves all show a brightness enhancement at optical
wavelengths of about 5\,percent lasting about 0.4 of an orbital cycle. The
effect shows little change in amplitude with colour. The amplitude of the
effect, the strength of the H$\alpha$ emission line and the variability of the
photometry of SDSS~1212+0136 all point to ongoing accretion in this binary at
a rate of at least $10^{-13} M_\odot/y$. The companion to the white dwarf is
close to filling its Roche lobe, which suggests that SDSS~1212+0136 may be a
normal polar in a low state. However, there are no recorded high-states for
this star.  The five epochs of photographic photometry in the USNO B-1 catalog
all show SDSS~1212+0136 at about 18th magnitude (Monet et al. 2003). Schmidt
et al. (2005a) have also discussed the possibility that SDSS~1212+0136 is a
polar in a low state but conclude that it is unlikely. A better estimate of
the mass transfer rate can be made using observations at X-ray wavelengths,
since the X-ray flux will be due entirely to accretion. 

There is reasonable
evidence for a smaller brightness increase during the orbital cycle,
approximately midway between large brightness enhancements. Given its orbital
phasing, amplitude, and possible increase with increasing wavelength, we
speculate that these smaller lightcurve bumps may be due to a reflection off
the companion. 

A better estimate of the mass of the companion will require a
radial velocity curve for the white dwarf. This will be difficult if the
Balmer lines are variable in profile since the semi-amplitude of the white
dwarf's spectroscopic orbit is expected to be only K$_{\rm WD} \approx 55$km/s.

\section*{\large \bf ACKNOWLEDGMENTS}
Exchange of ideas with Dr. Dave Kilkenny (SAAO) was helpful. 
PM thanks Dr Coel Hellier and Dr Boris G\"{a}nsicke
for discussing their opinions of the lightcurves.
Telescope time allocation by SAAO is gratefully acknowledged.

\pagebreak

{\bf Table 1.}\  \ The observing log: $T_{int}$ is the exposure time
and $N$ the number of useful measurements obtained during the run.
The last two columns give the nightly zeropoint offset from a ``check" star
of comparable brightness, and the standard deviation of the measurements
of the check star (see text for more details).\\\\

\begin{tabular}{ccccccc}
 Starting time &  Filter & $T_{int}$ & Run length & $N$ & $\Delta z$ &
$\sigma$ \\
      (HJD~2453800+)&  & (s)& (hours) &  & (mag) & (mmag)\\
& & & & & & \\
23.51882& $V$ &  80 & 1.6  & 50 & 0.435 & 15 \\
24.38333& $V$ & 100 & 4.4  &127 & 0.435 & 12 \\
25.45291& $V$ &  80 & 2.8  &101 & 0.447 & 14 \\
26.42660& $V$ &  80 & 2.7  &116 & 0.445 & 14 \\ 
& & & & & & \\
25.31673& $R$ & 100 & 3.2  & 94 &-0.086 & 10 \\
& & & & & & \\
27.38555& $-$ &  50 & 3.6  &178 & 0.025 & 8 \\
29.38165& $-$ &  50 & 4.4  &127 & 0.016 & 11 \\
51.29252& $-$ &  50 & 4.4  &221 &-0.003 & 9  
\end{tabular}

\vspace*{2cm}

{\bf Table 2.}\  \ Results of fitting the model of Eqn. (1) to the
various datasets. In the case of the $R$ band, $f_*$ from the white 
light dataset was used.\\\\

\begin{tabular}{ccccc}
 Filter & $f_*$ & 95\% conf. interval & $C_1$ & $C_2$ \\
        & (d$^{-1}$) & (d$^{-1}$) & (mmag) & (mmag) \\
 & & \\
 $V$ & 16.288  & (16.278, 16.298) & 17 & 9  \\
 $-$ & 16.2836  & (16.2827, 16.2844) & 16 & 9  \\
 $R$ &   &  & 14 & 8\\
 
\end{tabular}

\pagebreak

\begin{figure}
\epsfysize=8.0cm
\epsffile{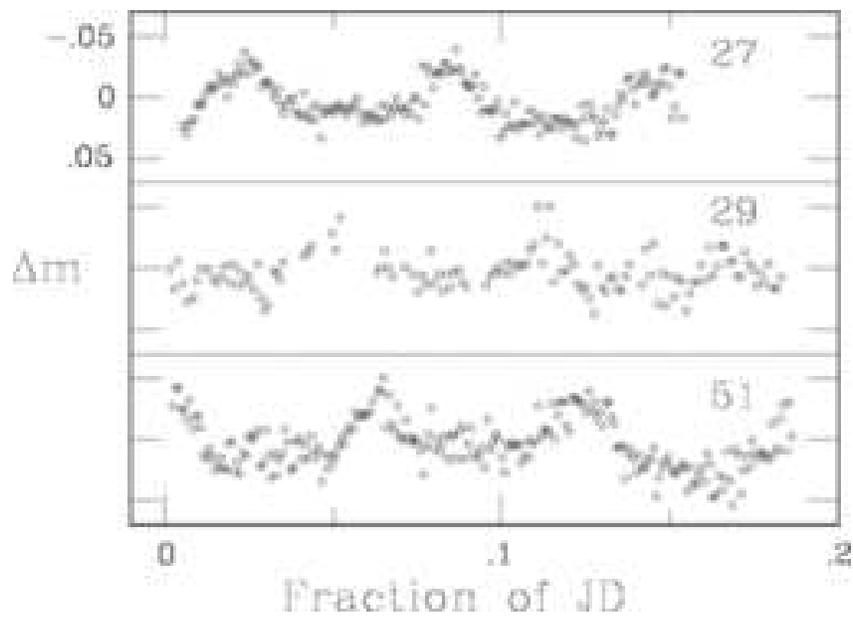}
\caption{Light curves obtained without any filter in the light beam.
Panels are labelled with the last two digits of the Julian Day of 
observation.}
\end{figure}

\begin{figure}
\epsfysize=11.0cm
\epsffile{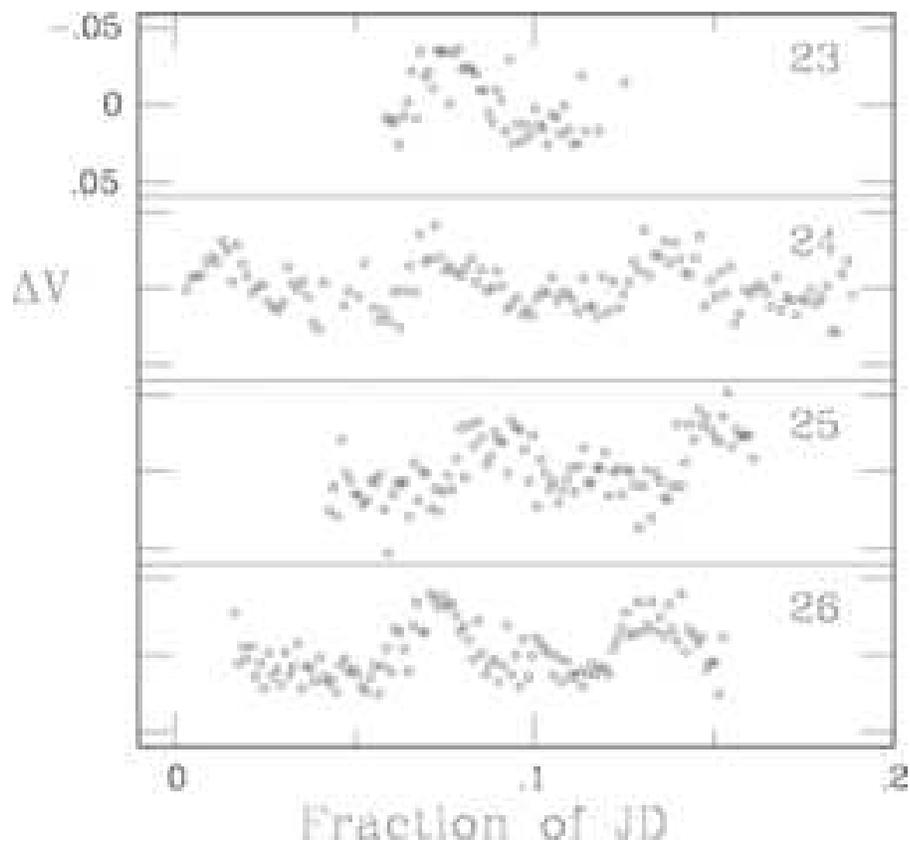}
\caption{Light curves obtained though the $V$ filter.
Panels are labelled with the last two digits of the Julian Day of 
observation.}
\end{figure}

\begin{figure}
\epsfysize=8.0cm
\epsffile{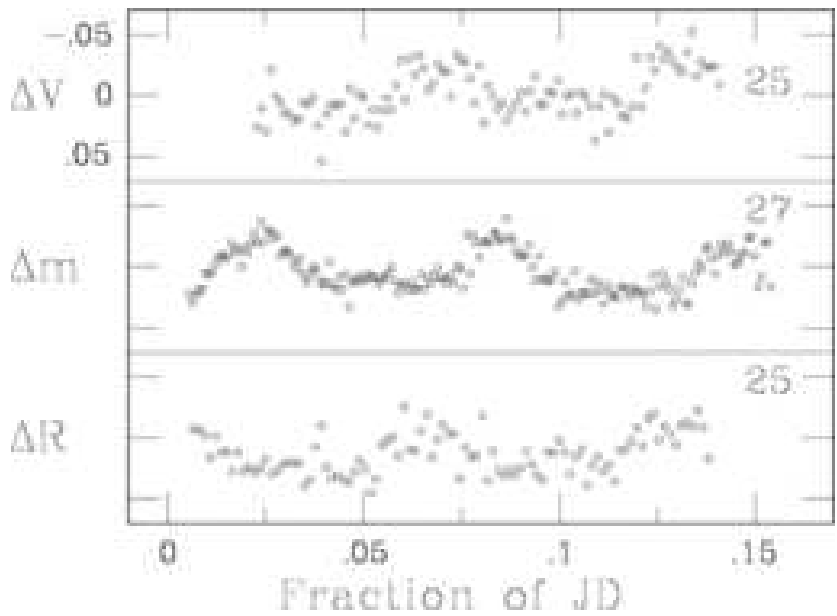}
\caption{A comparison of the $V$, white light, and $R$ lightcurves.
The scales on the three panels are the same.
Panels are labelled with the last two digits of the Julian Day of 
observation.}
\end{figure}

\begin{figure}
\epsfysize=8.0cm
\epsffile{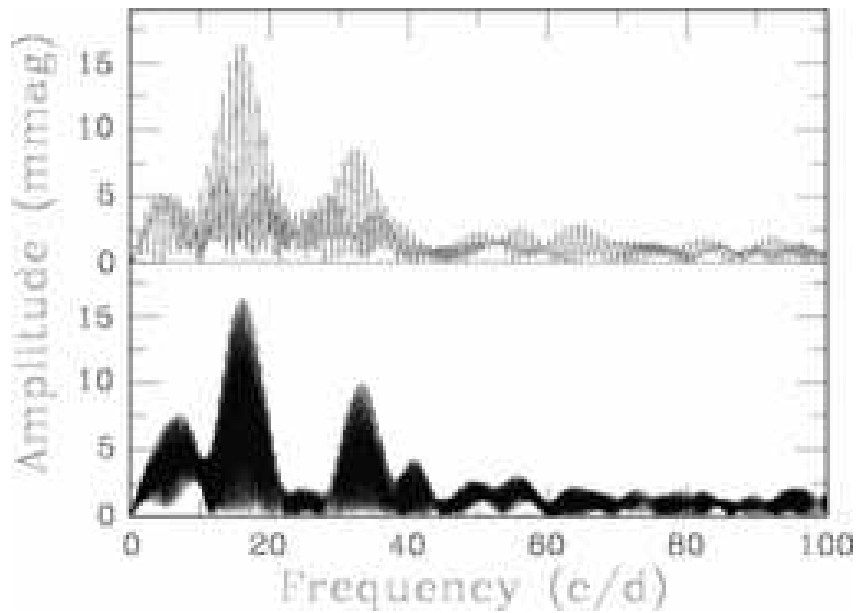}
\caption{Amplitude spectra of all the $V$ data (top panel) and all
the filter-less observations (bottom panel).}
\end{figure}

\begin{figure}
\epsfysize=8.0cm
\epsffile{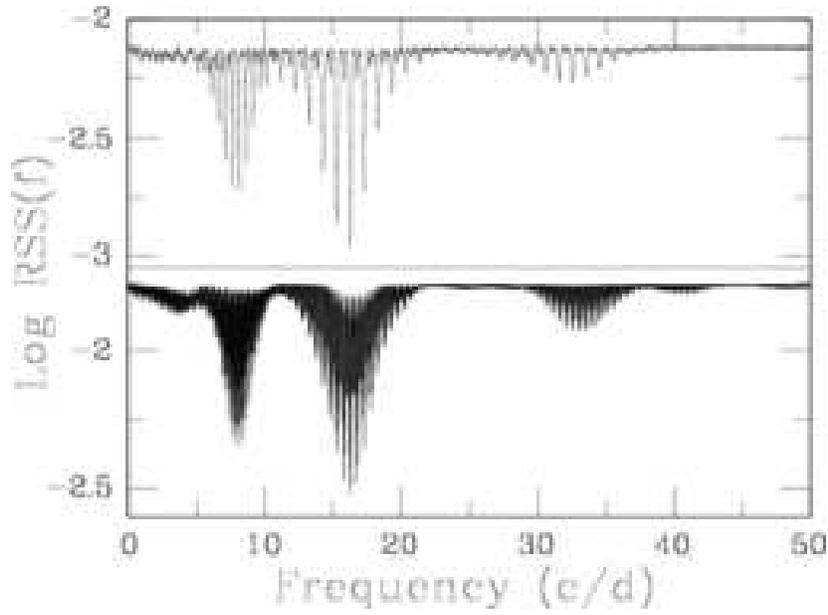}
\caption{The residual sum of squares for different trial frequency
fits to all the $V$ data (top panel)
and all the filter-less observations (bottom panel).}
\end{figure}

\begin{figure}
\epsfysize=8.0cm
\epsffile{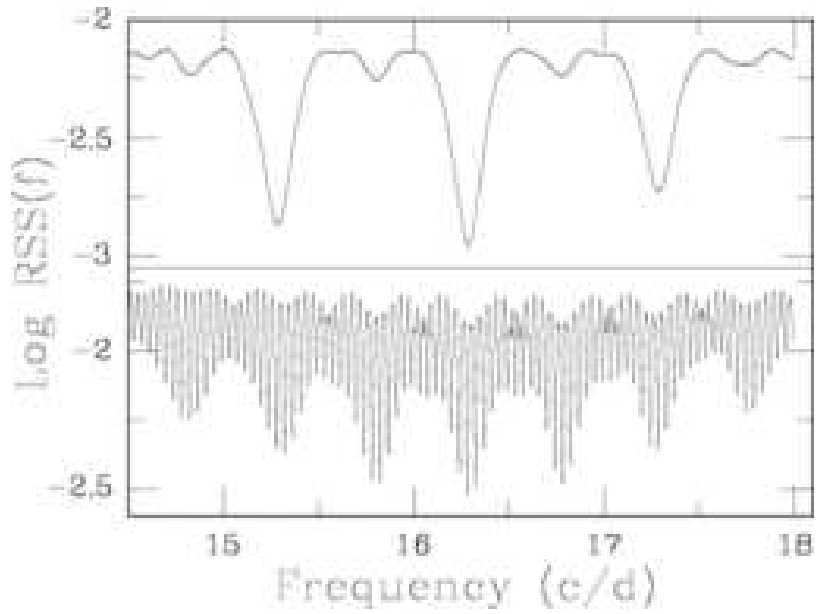}
\caption{Detail of the frequency interval of greatest interest in Fig. 5.
The top and bottom panels respectively show results for the $V$ filter
and for the white-light observations.}
\end{figure}

\begin{figure}
\epsfysize=8.0cm
\epsffile{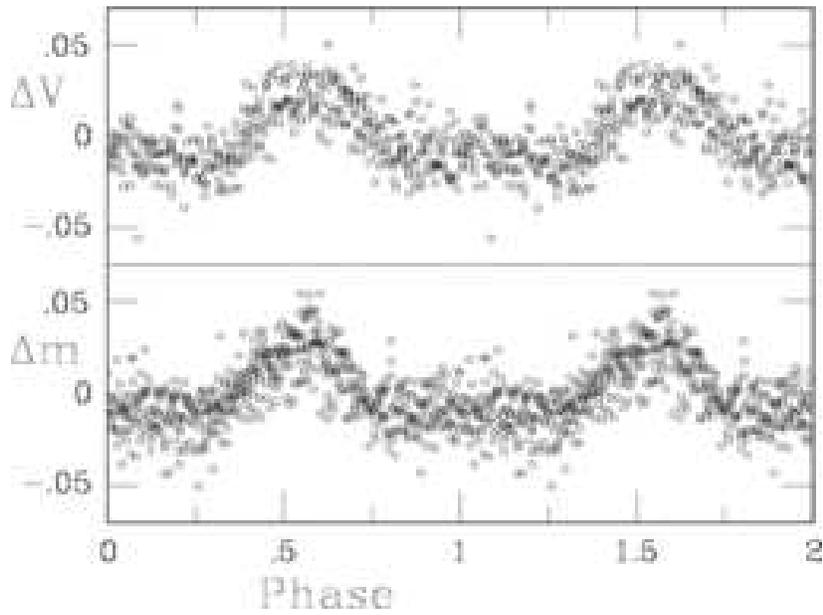}
\caption{All the $V$ (top panel) and white light (bottom panel) observations
folded with respect to the best-fitting period.}
\end{figure}

\begin{figure}
\epsfysize=8.0cm
\epsffile{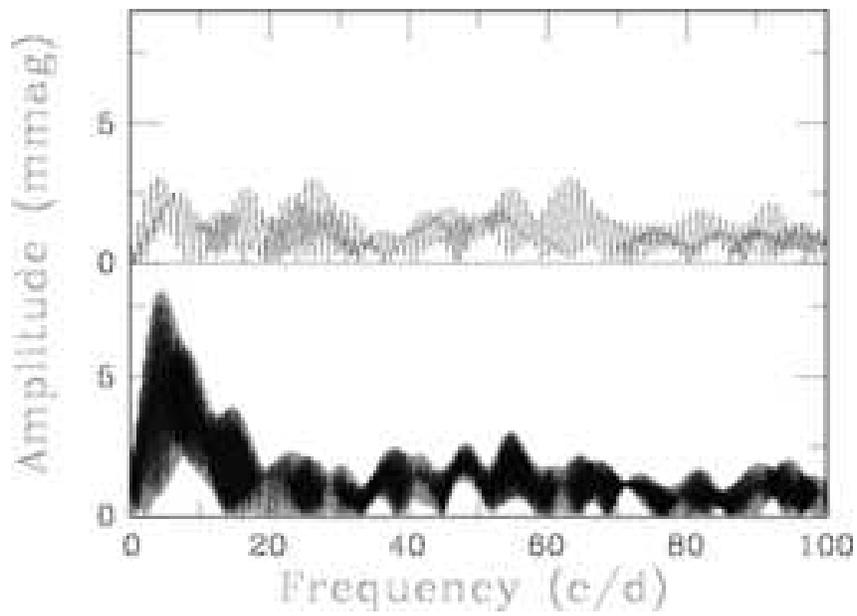}
\caption{Amplitude spectra of the residuals left after prewhitening by
the best-fitting two-term sinusoid. Results for $V$ are plotted in the
top panel, and results for white light in the bottom panel.}
\end{figure}

\begin{figure}
\epsfysize=10.0cm
\epsffile{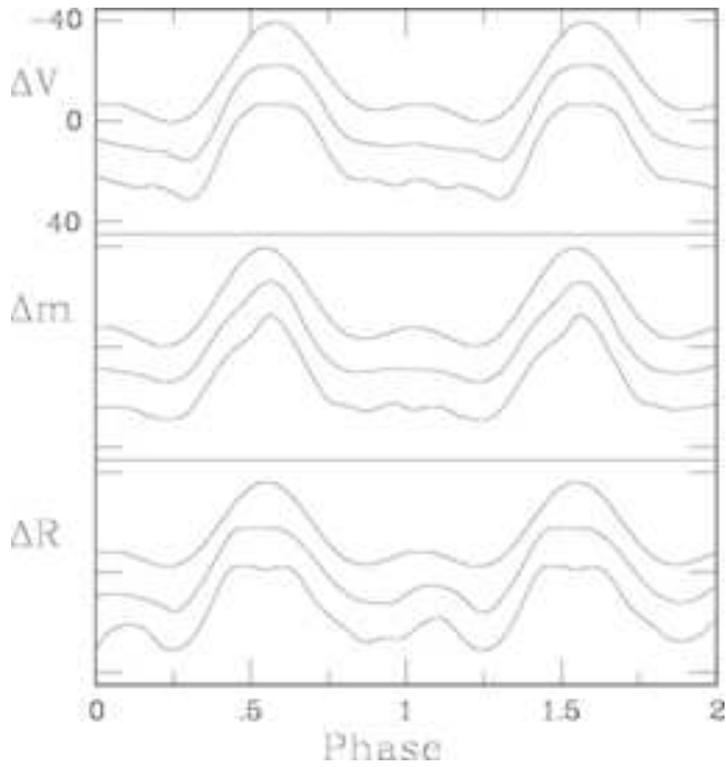}
\caption{Estimated light curve shapes for all the $V$-band (top panel), 
white-light (middle panel), and $R$-band (bottom panel) data respectively.
The zeropoints on the vertical axes are arbitrary, and the scales are in mmag. 
In each panel the top
curve is the shape described by Eqn. (1), while the other two curves are
non-parametric smooths of phase diagrams such as those in Fig. 7 -- see the
text for details.}
\end{figure}
\end{document}